\documentclass[12pt,draftcls,onecolumn]{IEEEtran}

\usepackage{hyperref}
\hypersetup{colorlinks,
citecolor=black,%
filecolor=black,%
linkcolor=black,%
urlcolor=black}
\usepackage{bm,mathptmx,amsfonts,amssymb,amstext,textcomp,amsmath}
\usepackage{times,verbatim,multirow,epsfig,graphics,graphicx,epstopdf}
\usepackage[usenames]{color}
\usepackage{tikz}

\DeclareMathAlphabet\mathbfcal{OMS}{cmsy}{b}{n}

\newcommand{\rea}{\mathbb{R}}

\newtheorem{mythm}{\bf Theorem}
\newtheorem{myass}{\bf Assumption}
\newtheorem{mylem}{\bf Lemma}
\newtheorem{myrem}{\bf Remark}

\ifCLASSINFOpdf
\else
\fi

\begin{document}

\title{Consensus-based control for a network of diffusion PDEs with boundary local interaction}

\author{Alessandro~Pilloni,~Alessandro~Pisano,~Yury~Orlov~and Elio~Usai
\thanks{A. Pilloni, A. Pisano and E. Usai are with the Department
of Electrical and Electronic Engineering (DIEE), University of Cagliari, Cagliari,
09123, Italy.
}
\thanks{Y. Orlov is with CICESE Research Center, Electronics and Telecommunication Department, Ensenada, Mexico.
}
\thanks{E-mail addresses: {$\lbrace$\tt\small alessandro.pilloni,pisano,eusai$\rbrace$ @diee.unica.it}, {\tt\small yorlov@cicese.mx}.
}
}



%

\maketitle

\begin{abstract}
In this paper the problem of driving the state of a network of identical agents, modeled by boundary-controlled heat equations, towards a common steady-state profile is addressed.
Decentralized consensus protocols are proposed to address two distinct problems. The first problem is that of steering the states of all agents towards the same constant steady-state profile which corresponds to the spatial average of the agents initial condition. A linear local interaction rule addressing this requirement is given.
The second problem deals with the case where the controlled boundaries of the agents dynamics are corrupted by additive persistent disturbances. To achieve synchronization between agents, while completely rejecting the effect of the boundary disturbances, a nonlinear sliding-mode based consensus protocol is proposed. Performance of the proposed local interaction rules are analyzed by applying a Lyapunov-based approach. Simulation results are presented to support the effectiveness of the proposed algorithms.

%
%
%
%
\end{abstract}

\begin{IEEEkeywords}
Average consensus, Synchronization, Heat equation, Boundary control, Sliding-mode control.
\end{IEEEkeywords}

\IEEEpeerreviewmaketitle

\section{Introduction}\label{sect1}

\IEEEPARstart{T}{}he problem of understanding when individual actions of interacting dynamical agents give rise to a coordinated collective behavior has received considerable attention in many research fields. Examples can be found, e.g., in system biology \cite{rosenfeld2013global}, sensors networks \cite{schenato2009average}, robotics \cite{ren2007consensus}, etc.

In the systems and control literature, the usual mathematical setup underlying this class of problems refers to a group of agents, each one described by a dynamical system with one or more inputs, along with a communication network. Agents connected by a communication link are said to be \emph{neighbors}, and can exchange information in either bidirectional or unidirectional manner. This raises the problem of designing \emph{decentralized} local interaction policies (where each agent can only access neighbors information) in order to orchestrate the global coordinated behavior of the network. Within this framework, the consensus problem seeks to enforce agreement amongst the states of networked dynamical systems by penalizing their local disagreement with the neighboring nodes in a dynamic manner. The reader should refer, e.g., to \cite{ren2007information,olfati2007consensus} for tutorial overviews of consensus-based control in the finite-dimensional setup.

There are deep connections between the consensus problem and certain Partial Differential Equations (PDEs), the diffusion equation in particular  \cite{sepulchre09}. For instance, discretizing in the spatial domain the one dimensional diffusion equation yields a high-dimensional system of networked first-order continuous-time integrators interacting through a linear Laplacian-based consensus protocol. More generally, the application of finite-difference approximations of PDEs results in the spatial variables being mapped into the agent indexes, and the spatial derivatives being transformed into links between neighbors.
Owing on the  deep connections between the consensus problem and certain Partial Differential Equations (PDEs) (see e.g. \cite{sepulchre09}), some authors (see e.g. \cite{advectiongraps,Advection-Diffusion}) have exploited (discretized forms of) several distributed parameter systems, such as advection and diffusion-advection equations, to derive more effective consensus protocols with improved convergence features. In spite of this intimate relationship between consensus algorithms and certain discretized PDEs, the consensus problem for a network of agents modeled as distributed parameter systems has not received yet the same level of attention than its finite-dimensional counterpart.


The following papers \cite{chao2007consensus,tricaud2009optimal,demetriou2010design,demetriou2009natural,demetriou2010guidance,demetriou2012enforcing,DeploymentPDETAC2015,demetriou2013synchronization}, which have investigated different aspects of consensus and synchronization in the distributed parameter systems setting, are worth to mention.
In \cite{chao2007consensus,tricaud2009optimal}, application of consensus to controlling mobile actuators in diffusion processes is discussed. The problem of designing consensus filters for state estimation in order to effectively integrate local information coming from a distributed spatial domain has been widely addressed \cite{demetriou2010design,demetriou2010guidance,demetriou2012enforcing}, both in the non-adaptive and adaptive setting. In \cite{li2014exact}, exact synchronization was achieved for a set of coupled wave processes, each one equipped with a boundary control input.

In \cite{PdEFerrari,bliman2008average}, the authors present consensus algorithms in the framework of Multi-Agent Systems (MASs) modeled by continuous-time Partial difference Equations (PdEs) on graphs. Conceptually, this class of PdEs mimics PDEs in spatial domains having a graph structure, and many mathematical tools of functional analysis for studying PdEs are completely analogous to the ones developed for PDEs. In \cite{MRACNAHS2010,MRACNAHS2008}, PdE-based model reference adaptive control laws are designed for a network of mobile agents to track desired deployment trajectories .

Within a related framework, the recent work \cite{DeploymentPDETAC2015} studies the 3D agents deployment problem by PDE techniques, treating the agents as a continuum and modeling their interaction through a complex-valued diffusion-reaction PDE in a 2D spatial domain. An explicit backstepping-based local boundary control is designed to stabilize a variety of open-loop unstable deployment manifolds. \cite{MeuKrs2011} presents a feedforward controller for multi-agent deployment by using a flatness-based motion planning method for PDEs. Reaction-advection-diffusion PDEs are used in  \cite{DeploKrstTACC11} along with a backstepping design for leader-enabled agents deployment onto planar curves. Similarly, hyperbolic PDE models are used to design decentralized control laws for large vehicular platoons \cite{BaroahTAC2009} and to analyze networks of oscillators \cite{MauroySep2013}.

In \cite{demetriou2013synchronization}, synchronization and consensus problems have been studied for a network of agents modeled by a class of parabolic PDEs and communicating through undirected communication topologies. In \cite{demetriou2013synchronization}, which appears to be more closely related to the present investigation among the existing references, some noticeable results are attained. First, the case of all-to-all communication between a set of identical agents is investigated and then the more realistic communication topology where each agents can only communicate with a limited set of neighboring agents is studied. In addition, a general abstract formulation of the underlying agents dynamics is introduced, and linear consensus controllers are presented to ensure agents agreement with a guaranteed convergence rate.
It should be noted however that 
the common steady-state profile of the agents is not established, and furthermore no perturbations are allowed to
affect the agents dynamics.

The present work aims to address consensus and synchronization problems for a network of dynamical agents, communicating through an undirected and connected static topology, provided that agents dynamics are governed by a class of diffusion PDEs with Neumann-type boundary actuation. The contribution of the paper is twofold. Firstly, a linear local interaction strategy whose implementation requires collocated boundary sensing only is proposed. With this strategy, it is shown that the agents states eventually converge $\lq\lq$pointwise-in-space" towards a common constant distribution whose value is given by the spatial average of the agents' initial conditions. Thus, the well-known average consensus algorithm is generalized from a network of integrators to the infinite-dimensional setting of networked heat processes.

{ Secondly, the more complex scenario where the agents dynamics are perturbed by a class of boundary disturbances, is considered. Based on the second-order sliding-mode control approach \cite{pisanousaisurvey}, a nonlinear protocol is developed to extend the results of \cite{pilloni2013twisting} from a network of double integrators to the infinite-dimensional framework of networked PDEs. A dynamic input extension, similar to that presented in \cite{pisano2012boundary} for stabilizing a unique perturbed diffusion PDE, results in continuous boundary control actions thereby alleviating chattering and yielding another step beyond \cite{pilloni2013twisting}. It is demonstrated that the proposed nonlinear local interaction protocol, which only employs sensors located at the controlled boundaries, enforces the asymptotic synchronization between the agents states while rejecting the persistent matching boundary perturbations.} 

{  The motivation to the present investigation comes, e.g., from networked systems of perturbed heat equations that can occur in modeling and controlling industrial furnaces. Heating of certain industrial furnaces (see, e.g., \cite[Sect. 1.A]{caponettecc}) is made through electrically heated bars aiming to enforce a uniform temperature distribution inside the furnace. 
Considering these bars as a network of heaters and applying collaborative consensus-based synthesis might be useful in improving the overall performance of the furnaces.
Exploiting the present results in specific application domains certainly requires additional work which is beyond the scope of the present paper.}


The paper is organized as follows. In Section~\ref{sect2} some mathematical preliminaries and useful properties and definitions are recalled. The linear average consensus algorithm 
is presented in Section~\ref{sect3} whereas the nonlinear algorithm, providing  robust synchronization in the presence of boundary perturbations, is described in  Section~\ref{sect4}. Simulation results, supporting the proposed designs, are given in Section~\ref{sect5}, and conclusions and perspectives for next investigations are collected in the final Section~\ref{sect6}.

\section{Mathematical Preliminaries and Notations}\label{sect2}

\subsection{Useful definitions and properties}

The $l_p$-norm and the $l_\infty$-norm of the real-valued N-dimensional vector ${x} = [x_1,\dots,x_N]^T \in \rea^N$ are defined as $\|{x}\|_{\mathrm{{p}}}=\left( \sum_{i=1}^N |x_i|^\mathrm{p}\right)^{1/\mathrm{p}}, \; 1\leq \bm{\mathrm{p}} < \infty$ and $\|{x}\|_\infty=\max_{1 \leq i \leq N}\left\lbrace|x_1|,|x_2|,\dots,|x_N|\right\rbrace$.
For the $l_1$- and $l_2$- norms, the following inequality holds \cite{khalil2002nonlinear}:
\begin{gather}
\|{x}\|_{ 2} ~\leq ~ \|{x}\|_{ 1} ~\leq  ~\sqrt{N} ~\|{x}\|_{ 2}.\label{app1:prop1}
\end{gather}

Let $\mathrm{p},\mathrm{q}\geq 1$ be given, such that ${1}/{\mathrm{p}}+{1}/{\mathrm{q}}=1$. Then the next chain of inequalities is in force \cite{khalil2002nonlinear}
\begin{equation}{\small
\left|{x}^T{y}\right|\leq\|{x}\|_{\mathrm{p}}\|{y}\|_\mathrm{q} \leq \frac{\|{x}\|_{\mathrm{p}}^\mathrm{p}}{\mathrm{p}}+\frac{\|{y}\|_{\mathrm{q}}^\mathrm{q}}{\mathrm{q}}.}
\label{app1:HolderInequality}
\end{equation}

Operator $\mathrm{sign}(v)$, $v \in \rea$, stands for the multi-valued function
\begin{equation}
\mathrm{sign}\left(v\right)\in
\left\lbrace
\begin{array}{cl}
1 & \quad\mathrm{if}\quad  v>0\\
\left[-1, 1\right] & \quad\mathrm{if}\quad  v=0\\
-1 & \quad \mathrm{if}\quad v<0\\
\end{array}
\right.,
\label{sect1:SIGN}
\end{equation}
whereas ${\mathrm{Sign}}\left({x}\right)$ stands for the vector ${\mathrm{Sign}}\left({x}\right)=
\left[ \mathrm{sign}\left(x_1\right), \mathrm{sign}\left(x_2\right),\dots,\mathrm{sign}\left(x_N\right)\right]^T$.

The identity matrix of dimension $N$ is denoted as ${\mathbfcal I}_{N\times N}\in\rea^{N\times N}$, whereas ${1}_N=[1,1,\dots,1]^T\in\rea^N$ and ${0}_N=[0,0,\dots,0]^T\in\rea^N$ stand for
the all-ones and all-zeros vectors.

\vspace{0.1cm}

$\mathrm{H}^r(0,1)$, with $r=0,1,2,\dots$, denotes the Sobolev space of absolutely continuous scalar functions $z(\varsigma)$ on the domain $(0,1)$, with square integrable derivatives $z^{(k)}(\varsigma)$ up to order $\ell$ and the $\mathrm{H}^r$-norm $\|{z}(\cdot)\|_{\mathrm{H}^r(0,1)}
=\sqrt{\int_0^1 \sum_{k=0}^r \left[{z}^{(k)}(\xi)\right]^2 ~d\xi}.$

Then, the notations
\begin{equation}
\mathrm{H}^{r,N}(0,1) =
\begin{array}{c}
\underbrace{\mathrm{H}^r(0,1) \times \mathrm{H}^r(0,1) \times \ldots \times \mathrm{H}^r(0,1)} \\
N \;\; \texttt{times}
\end{array}
\end{equation}
and
\begin{eqnarray}\label{app1:L2normVector}
  \|{w}(\cdot)\|_{\mathrm{H}^{r,N}(0,1)} = \sqrt{\sum_{i=1}^{N}   \|w_i(\cdot)\|^2_{\mathrm{H}^r(0,1)}}
\end{eqnarray}
for the corresponding norm of the vector ${w}(\varsigma)=[w_1(\varsigma),\dots,w_N(\varsigma)]^T\in H^{r,N}(0,1)$ are utilized. The simplified notation $\|{z}(\cdot)\|_{\mathrm{H}^r}=\|{z}(\cdot)\|_{\mathrm{H}^r(0,1)}$, $\|{w}(\cdot)\|_{\mathrm{H}^{r,N}} = \|{w}(\cdot)\|_{\mathrm{H}^{r,N}(0,1)}$ will be adopted throughout.

For later use, an instrumental lemma is further presented:
\vspace{0.15cm}
\begin{mylem}\label{lemma1:pisanoResult}{\it
Let ${b}(\varsigma)\in\mathrm{H}^{1,N}(0,1)$. Then, the following inequality holds:
\begin{equation}
\|{b}(\cdot)\|_{\mathrm{H}^{0,N}}^2\leq 2\left(\|{b}(i)\|_{{2}}^2+\|{b}_\varsigma(\cdot)\|_{\mathrm{H}^{0,N}}^2\right),\;\; i=0,1\nonumber
\label{app1:pisanoResult}
\end{equation}}
\end{mylem}
\begin*{\bf Proof of Lemma \ref{lemma1:pisanoResult}:}
It was proven \cite[Lemma 1]{pisano2012boundary}  that, with reference to a generic scalar function $z(\varsigma) \in H^1(0,1)$, the next estimate holds:
\begin{equation}
 \|z(\cdot)\|_{\mathrm{H}^0}^2\leq 2(z(i)^2+\|z_\varsigma(\cdot)\|_{\mathrm{H}^0}^2), \;\; i=0,1.
\label{lm1autom}
\end{equation}

Now let ${b}(\varsigma)=[b_1(\varsigma),b_2(\varsigma),\dots,b_N(\varsigma)]^T$ and ${b}_\varsigma(\varsigma)=[b_{\varsigma 1}(\varsigma),b_{\varsigma 2}(\varsigma),\dots,b_{\varsigma N}(\varsigma)]^T$ where $b_k(\varsigma)\in\mathrm{H}^1(0,1)$ $\forall k=1,2,\dots,N$. By applying definition \eqref{app1:L2normVector}, the following chain of relations is derived by virtue of \eqref{lm1autom} specified with $z(\cdot)=b_k(\cdot)$:
\begin{align}
\|{b}(\cdot)\|_{\mathrm{H}^{0,N}}^2&=\sum_{j=1}^N \|b_j(\cdot)\|_{\mathrm{H}^0}^2 \leq 2\sum_{j=1}^N\left(b_j(i)^2+\|b_{\varsigma j}(\cdot)\|_{\mathrm{H}^0}^2\right)=2\left(\|{b}(i)\|_{{2}}^2+\|{b}_\varsigma(\cdot)\|_{\mathrm{H}^{0,N}}^2\right), \;\; i=0,1.&&
\end{align}
Lemma~\ref{lemma1:pisanoResult} is proved.\hfill $\square$
\end*{}

\subsection{Algebraic Graph Theory definitions and properties}

We consider a set of $N$ dynamical agents along with an undirected static communication topology represented by the graph $\mathbfcal{G}(\mathbfcal{V},\mathbfcal{E})$, where $\mathbfcal{V}=\{1,\ldots,N\}$ is the set of vertices representing agents and $\mathbfcal{E}\subseteq \{\mathbfcal{V} \times \mathbfcal{V}\}$ is the set of edges representing the information flow among the agents. $\mathbfcal{N}_i=\{j \in \mathbfcal{V}:(i,j)\in \mathbfcal{E}\}$ denotes the set of neighbors of agent $i$. 
The topological structure of $\mathbfcal{G}$ is encoded in the so-called \emph{Laplacian Matrix} $\mathbfcal{L}=[\ell_{ij}]\in\rea^{N\times N}$ where
\begin{eqnarray}
\ell_{ij}:=\left\lbrace
\begin{array}{cl}
\left| \mathbfcal{N}_i\right| & \quad\mathrm{if}  \quad i=j\\
-1 & \quad \mathrm{if} \quad \left(i,j\right)\in\mathbfcal{E}\\
0 & \quad\mathrm{otherwise}\\
\end{array}
\right.
\label{sect1:Laplacian}
\end{eqnarray}
For undirected connected graphs, the matrix $\mathbfcal{L}$ is symmetric and positive semi-definite \cite{olfati2007consensus}, the properties
\begin{equation}\label{laplzs}
\mathbfcal{L}{1}_N=\mathbfcal{L}^T{1}_N=0_N
\end{equation}
hold by construction, and the corresponding eigenvalues $\lambda_i$, $i \in \mathbfcal{V} $, are such that $0=\lambda_1 < \lambda_2 \leq \ldots \leq \lambda_N$. The smallest nonzero eigenvalue $\lambda_2$ is known as \emph{algebraic connectivity} of $\mathbfcal{G}$.
Next lemma presents useful properties of vector norms involving the Laplacian matrix of the graph.

\begin{mylem}\label{lemma2:lam2}{\it For an undirected connected graph with Laplacian matrix $\mathbfcal{L}$, and with reference to any vector ${x} \in \rea^N$ such that ${1}_N^T {x} = 0$, the next relations are in force}
\begin{eqnarray}
 \lambda_N ||{x}||_2^2 \geq {x}^T \mathbfcal{L} {x} \geq \lambda_2 ||{x}||_2^2
\label{app1:QuadraticFormNorm}  \\
  \lambda^2_N ||{x}||_2^2 \geq \|\mathbfcal{L}{x}\|^2_{{2}} \geq \lambda^2_2 ||{x}||_2^2\label{sect1:laplacianPropertySQUARE}\\
 \|\mathbfcal{L}{x}\|_{{1}} \geq \lambda_2 \|{x}\|_{{2}} \label{sect1:laplacianProperty}
\end{eqnarray}
\end{mylem}
\begin*{\bf Proof of Lemma \ref{lemma2:lam2}:}
The left inequality in \eqref{app1:QuadraticFormNorm} comes from well-known properties of quadratic norms. The right inequality in \eqref{app1:QuadraticFormNorm} was proven in \cite[Th. 3]{olfati2007consensus}. To derive \eqref{sect1:laplacianPropertySQUARE}, observe that $\|\mathbfcal{L}{x}\|_{{2}}= \sqrt{{x}^T\mathbfcal{L}^2{x}}$. The eigenvalues $\left\{ 0, \lambda_2^2, \lambda_3^2, \ldots, \lambda_N^2\right\}$ of $\mathbfcal{L}^2$ are straightforwardly derived by squaring those of $\mathbfcal{L}$. Thus, the left inequality of \eqref{sect1:laplacianPropertySQUARE} follows from well-known properties of quadratic norms. Additionally, $\mathbfcal{L}^2 $ is symmetric and such that $\mathbfcal{L}^2 1_N = 0_N$, thus $1_N$ is the eigenvector associated to the zero eigenvalue of $\mathbfcal{L}^2 $. Therefore, the right inequality of \eqref{sect1:laplacianPropertySQUARE} follows from the Courant-Fisher Theorem that can be found, e.g., in \cite{horn1990matrix}. To reproduce \eqref{sect1:laplacianProperty}, it suffices to conclude from \eqref{app1:prop1} that   $\|\mathbfcal{L}\bm{x}\|_{{1}}\geq\|\mathbfcal{L}\bm{x}\|_{{2}}$ and then, by applying \eqref{sect1:laplacianPropertySQUARE}, to derive that $\|\mathbfcal{L}\bm{x}\|_{{2}} \geq \lambda_2 ||\bm{x}||_2$. Lemma~\ref{lemma2:lam2} is proved.\hfill $\square$
\end*{}

\section{Average consensus for networked heat processes}\label{sect3}


A network of $N$ dynamical agents whose communication topology is described by an undirected connected static graph $\mathbfcal{G}(\mathbfcal{V},\mathbfcal{E})$ is under study. The $i$-th agent has state $Q_i(\varsigma,t)$, $i\in\mathbfcal{V}$, with the spatial variable $\varsigma\in(0,1)$ and time variable $t \geq 0$. Let $Q(\varsigma,t)=\begin{bmatrix}Q_1(\varsigma,t),Q_2(\varsigma,t),\dots,Q_N(\varsigma,t)\end{bmatrix}^T$  be the vector collecting the states of all agents, and let the dynamics of $Q(\varsigma,t)$ be governed by the vector heat equation
\begin{equation}
Q_t(\varsigma,t)=\theta\cdot Q_{\varsigma\varsigma}(\varsigma,t),
\label{sect1:barDynamicAVE}
\end{equation}
The scalar parameter $\theta \in \rea^+$ is a positive unknown coefficient, called $\lq\lq$diffusivity parameter", which is supposed to be identical for all agents. Throughout, Neumann-type Boundary Conditions (BCs) of the form
\begin{equation}
Q_\varsigma(0,t)=0,\quad\quad Q_\varsigma(1,t)=U(t),
\label{sect1:barBCsAVE}
\end{equation}
are considered, where $U(t)=\begin{bmatrix}u_1(t),u_2(t),\dots,u_N(t)\end{bmatrix}^T\in\rea^N$ is a modifiable source term (boundary vector control input).

The Initial Conditions (ICs) are
\begin{equation}
Q(\varsigma,0)= Q_0(\varsigma) 
\label{sect1:barICsAVE}
\end{equation}

{ To deal with  classical solutions of class $\mathrm{H}^{2,N}(0,1)$, the admissible initial functions are specified by the next assumption.}

\vspace{0.1cm}
\begin{myass}\label{assumption:1_distrurbanceAVE}
{\it
{ The initial function $Q_0(\varsigma)$ in the ICs \eqref{sect1:barICsAVE} is assumed to be of class $\mathrm{H}^{2,N}(0,1)$} and compatible to the BCs $Q_{0\varsigma}(0)=0$ and $ Q_{0\varsigma}(1)=U(0)$.
}\end{myass}

\vspace{0.1cm}


\vspace{0.1cm}

The objective of the present section is to introduce a linear local interaction strategy providing closed-loop stability and the point-wise consensus condition
\begin{equation}
 \lim\limits_{t\rightarrow\infty} Q(\varsigma,t) = Q^* \cdot 1_N,~~ \forall \varsigma \in (0,1),
\label{sect1:consensusDefinitionAVE2}
\end{equation}
where the constant
\begin{equation}
 Q^* = \frac{1}{N} \int_0^1 1_N^T Q_{0}(\varsigma) d\varsigma
\label{sect1:consensusDefinitionAVE3}
\end{equation}
corresponds to the spatial averaging of the agents initial conditions.

To achieve the control goal, the local interaction protocol
\begin{equation}
{U}\left( t \right)=- \mathbfcal{L}Q(1,t)
\label{sect1:consensusProtocol0AVE}
\end{equation}
is proposed.  Under the assumptions, imposed on the ICs and BCs, the well-posedness of the system in question is straightforwardly verified by applying \cite[Theorem 2.1.10]{curtain1995introduction} to the classical solutions of the homogeneous linear Boundary-Value Problem (BVP) \eqref{sect1:barDynamicAVE}-\eqref{sect1:barICsAVE}, \eqref{sect1:consensusProtocol0AVE}.

We are now in a position to state the first main result of this paper.

\begin{mythm}\label{Theorem1AVE}
Consider the multi-agent system \eqref{sect1:barDynamicAVE}-\eqref{sect1:barICsAVE},  with Assumption \ref{assumption:1_distrurbanceAVE}, communicating through an undirected connected static graph with Laplacian matrix $\mathbfcal{L}$. Let it be subject to the boundary local interaction control strategy \eqref{sect1:consensusProtocol0AVE}.
Then, the closed-loop system is stable in the space $H^2(0,1)$ and the average consensus condition  \eqref{sect1:consensusDefinitionAVE2}-\eqref{sect1:consensusDefinitionAVE3} is achieved. {\hfill $\square$}
\end{mythm}

\vspace{0.2cm}
\begin*{\bf Proof of Theorem \ref{Theorem1AVE}:}\label{proofTheorem1} {  The stability of the closed-loop BVP is established by involving the Lyapunov function
\begin{flalign}
V_1(t)=&\frac{1}{2}\int_0^1 Q^T(\xi,t)Q(\xi,t)d\xi
\label{sect2:V1defff}
\end{flalign}
whose time derivative, estimated along the solutions of the BVP \eqref{sect1:barDynamicAVE}-\eqref{sect1:barICsAVE},  \eqref{sect1:consensusProtocol0AVE}, is non-positive definite:
\begin{flalign}
\dot V_1(t)=\theta\int_0^1 Q^T(\xi,t)Q_{\xi\xi}(\xi,t)d\xi = -\theta \|Q_{\varsigma}(\cdot,t)\|_{\mathrm{H}^{0,N}}^2 -\theta Q(1,t)^T \mathbfcal{L}Q(1,t)\leq -\theta \|Q_{\varsigma}(\cdot,t)\|_{\mathrm{H}^{0,N}}^2.  \label{se20}
\end{flalign}

To get \eqref{se20} integration by parts, and BCs  \eqref{sect1:barBCsAVE},\eqref{sect1:consensusProtocol0AVE}, and the semi-definite positiveness of the Laplacian matrix   $\mathbfcal{L}$ were utilized. Relation \eqref{se20} ensures that the system is stable in the space $H^0(0,1)$. Since the BVP is linear, and the ICs \eqref{sect1:barICsAVE} are of class $H^2(0,1)$, the stability remains in force in the space $H^2(0,1)$.

Next, let us note that the eigenspace of the closed-loop BVP \eqref{sect1:barDynamicAVE}-\eqref{sect1:barICsAVE},  \eqref{sect1:consensusProtocol0AVE}, associated with the zero eigenvalue, is one dimensional and it is spanned by the uniform distribution $Q(\varsigma)=  {1}_N $. In turn, all the remaining eigenvalues are strictly negative  because the BVP \eqref{sect1:barDynamicAVE}-\eqref{sect1:barICsAVE},  \eqref{sect1:consensusProtocol0AVE} has been shown to be stable.

Furthermore, one observes that the projection
\begin{flalign}
\frac{1}{N} \int_0^1 {1}^T_N Q(\xi,t)d\xi 1_N
\end{flalign}
of the solution of the closed-loop system to the eigenspace, associated to the zero eigenvalue, remains constant, whereas all the remaining modes tend to zero because the other eigenvalues are strictly negative. It follows that the state $Q(\varsigma,t)$ eventually converges point-wise to the constant spatial distribution
\begin{flalign}
\frac{1}{N} \int_0^1 {1}^T_N Q_0(\xi)d\xi 1_N = Q^* \cdot 1_N
\end{flalign}
thereby establishing relations \eqref{sect1:consensusDefinitionAVE2}-\eqref{sect1:consensusDefinitionAVE3}. This completes the proof of Theorem~\ref{Theorem1AVE}.} {\hfill $\square$}.

%
%

\end*{}


\section{Robust synchronization for networked heat processes with perturbations}\label{sect4}


A perturbed version of the BVP \eqref{sect1:barDynamicAVE}-\eqref{sect1:barICsAVE}, with the only difference in the BCs \eqref{sect1:barBCsAVE} which now take the perturbed form
\begin{equation}
Q_\varsigma(0,t)=0, \quad Q_\varsigma(1,t)=U(t)+\Psi(t),
\label{sect1:barBCs}
\end{equation}
is under investigation, where $\Psi(t)=\begin{bmatrix}\psi_1(t),\psi_2(t),\dots,\psi_N(t)\end{bmatrix}^T\in\rea^N$ represents an uncertain, sufficiently smooth, persistent disturbance.

The class of admissible ICs and disturbances is specified by the next assumption.
\vspace{0.15cm}
\begin{myass}\label{assumption:1_distrurbance}
{\it
The initial function $Q_0(\varsigma)$ 
{ is assumed to be of class $\mathrm{H}^{4,N}(0,1)$} and compatible to the  perturbed BCs $Q_{0\varsigma}(0)=0,\; Q_{0\varsigma}(1)=\Psi(0)$,  whereas the disturbance $\Psi(t)$ is supposed to be twice continuously differentiable, and there exists an \emph{a-priori} known constant $\Pi > 0$ such that
\begin{equation}
\left\|\dot{\Psi}(t)\right\|_\infty \leq~\Pi.
\label{sect1:disturbanceCondition}
\end{equation}
}\end{myass}

\vspace{0.05cm}

{ Note that for technical reasons (see Remark \ref{remark:2orderConsensus} below) higher degree of smoothness of the ICs is required in the present perturbed scenario.} The objective of the present section is to develop a local interaction strategy providing the attainment of the synchronization condition
\begin{equation}
\lim\limits_{t\rightarrow\infty} \left|Q_i(\varsigma,t)-Q_j(\varsigma,t)\right|=0,~~\forall~i,j\in \mathbfcal{V}, \forall \varsigma \in (0,1),
\label{sect1:consensusDefinitionAVE}
\end{equation}
despite the presence of the uncertain boundary disturbance $\Psi(t)$ of arbitrary shape and possibly unbounded in magnitude.

To achieve the control goal, 
the following dynamic local interaction protocol
\begin{equation}
\dot{U}\left( t \right)=\dot{U}_1(t)+\dot U_2(t)
\label{sect1:consensusProtocol0}
\end{equation}
is proposed, with
\begin{flalign}
\dot{U}_1(t)=&-a \mathrm{Sign}\left(\mathbfcal{L}Q(1,t)\right)
-b \mathrm{Sign}\left(\mathbfcal{L}Q_t(1,t)\right)\label{sect1:consensusProtocol1}\\
\dot{U}_2(t)=&-W_1\cdot\mathbfcal{L}Q(1,t)-W_2\cdot\mathbfcal{L}Q_t(1,t)-W_3 \cdot Q_t(1,t)\label{sect1:consensusProtocol2} \\ {U}_1(0)=&{U}_2(0)= 0_N.
\label{sect1:contricss}
\end{flalign}
The initial values $U_1(0), U_2(0)$ are all set to zero to verify the compatibility\footnote{See, e.g., \cite{vaz} for the need of certain
compatibility conditions in the dynamic boundary control synthesis.} $Q_{0\varsigma}(1)=U(0)+\Psi(0)$ to the BCs \eqref{sect1:barBCs} at $\varsigma=1$. In \eqref{sect1:consensusProtocol1}-\eqref{sect1:consensusProtocol2},
$a$, $b$, $W_1$, $W_2$ and $W_3$ are nonnegative tuning constants subject to certain design inequalities that will be constructively derived in the sequel. 

It is worth to note that the discontinuities affect the time derivative of the boundary control vector, whereas the boundary control signal is smoothed by passing these discontinuities through an integrator, thereby alleviating chattering.

\begin{myrem}\label{remark:2orderConsensus}
Although  the state derivative is normally not permitted to be employed in the synthesis (as it generally induces algebraic loops), its use becomes acceptable when dynamic input extension is performed, what is indeed the case of the present dynamic synthesis where the input signal passes through an integrator. By virtue
of this, the system state is augmented by $Q_t$ being viewed as a component of the augmented state vector $(Q,Q_t)\in \mathrm{H}^{4,N}(0,1) \times \mathrm{H}^{2,N}(0,1) $  which is particularly why the initial function $Q_0(\varsigma)$ was assumed to be of class $ \mathrm{H}^{4,N}(0,1)$.  {\hfill $\square$}

\end{myrem}
\vspace{0.15cm}
The well-posedness of the underlying closed-loop system, under the assumptions, imposed on the ICs and BCs, is actually verifiable in accordance with \cite[Theorem 3.3.3]{curtain1995introduction} by taking into account that the dynamic local interaction rule \eqref{sect1:consensusProtocol0}-\eqref{sect1:consensusProtocol2} is  twice piece-wise continuously differentiable along the state trajectories. Thus, in the remainder, it is assumed the following:

\begin{myass}\label{ass2}{\it
The closed loop networked system  \eqref{sect1:barDynamicAVE}-\eqref{sect1:barICsAVE}, \eqref{sect1:barBCs}, \eqref{sect1:consensusProtocol0}-\eqref{sect1:contricss} possesses a unique Filippov solution $Q(\cdot,t)\in
\mathrm{H}^{4,N}(0,1)$ and its time derivative $Z(\cdot,t) = Q_{t}(\cdot,t)\in \mathrm{H}^{2,N}(0,1)$ verifies the auxiliary boundary-value problem
\begin{align}
Z_{t}(\varsigma,t)&=\theta  Z_{\varsigma\varsigma}(\varsigma,t) && \label{sect1:NetworkDynamic0}\\
Z_{\varsigma}(0,t)& =0, \quad \quad
Z_{\varsigma}(1,t)=\dot{U}(t)+\dot{\Psi}(t), &&
\label{sect1:NetworkDynamic1}\\
\quad Z(\varsigma,0)&=\theta  Q_{0\varsigma\varsigma}(\varsigma)\in \mathrm{H}^{2,N}(0,1). && \label{sect1:NetworkDynamic2}
\end{align}}
\end{myass}

\vspace{0.15cm}

Extension of the Filippov solution concept towards the infinite dimensional setting can be found, e.g., in \cite{orlov2008discontinuous}. Notice that \eqref{sect1:NetworkDynamic0}-\eqref{sect1:NetworkDynamic1} are formally obtained by differentiating \eqref{sect1:barDynamicAVE}-\eqref{sect1:barICsAVE}, \eqref{sect1:barBCs}, in the time variable $t$, whereas the IC \eqref{sect1:NetworkDynamic2} is straightforwardly derived from \eqref{sect1:barDynamicAVE} and \eqref{sect1:barICsAVE}.



It is customary \cite{olfati2007consensus} to formalize the achievement of consensus through the annihilation of appropriate (N-dimensional) $\lq\lq$disagreement" vectors. Generalizing \cite{demetriou2013synchronization}, the following distributed disagreement vectors
\begin{eqnarray}
{\delta}_1(\cdot,t)&=&[\delta_{11}(\cdot,t), \ldots,\delta_{1N}(\cdot,t)]^T = \mathbfcal{L}_\mathbfcal{C}\ Q(\cdot,t),  \label{sect2:DisagreementVectorAVE} \\
{\delta}_2(\cdot,t)&=&[\delta_{21}(\cdot,t), \ldots,\delta_{2N}(\cdot,t)]^T={\delta}_{1t}(\cdot,t)=
\mathbfcal{L}_\mathbfcal{C}\ Q_t(\cdot,t),  \label{sect2:DisagreementVector2}\\
\mathbfcal{L}_\mathbfcal{C} &=& \left(\mathbfcal{I}_{N\times N}-\frac{\bm{1}_N\cdot\bm{1}_N^T}{N} \right),\label{LcAVE}
\end{eqnarray}
will be considered in the present investigation for analysis purposes. 
The next properties hold due to \eqref{laplzs} and \eqref{sect2:DisagreementVectorAVE}-\eqref{LcAVE}
\begin{align}
{1}^T_N {\delta}_{1}(\varsigma,t)&= {1}^T_N {\delta}_{2}(\varsigma,t)=0, \quad \forall \varsigma \in [0,1], &&\label{sect2:1tlamn1AVE} \\
\mathbfcal{L} \mathbfcal{L}_\mathbfcal{C}&=\mathbfcal{L}_\mathbfcal{C} \mathbfcal{L}=\mathbfcal{L} &&\label{LcLAVE}
\end{align}
thereby implying that
\begin{align}
&\mathbfcal{L} {\delta}_1(\varsigma,t)= \mathbfcal{L} \mathbfcal{L}_\mathbfcal{C} Q(\varsigma,t)= \mathbfcal{L} Q(\varsigma,t), && \label{sect2:LLCDIAAVE} \\
 &\mathbfcal{L} {\delta}_2(\varsigma,t)= \mathbfcal{L} \mathbfcal{L}_\mathbfcal{C} Q_t(\varsigma,t)= \mathbfcal{L} Q_t(\varsigma,t) &&  \label{sect2:LLCDIAt}
\end{align}

The BVP governing the dynamics of the disagreement vectors now reads as
\begin{align}
&\begin{array}{l}
{\delta}_{1t}(\varsigma,t)=\delta_2(\varsigma,t),\\
{\delta}_{2t}(\varsigma,t)=\theta \delta_{2\varsigma\varsigma}(\varsigma,t),
\end{array}\label{sect2:disDynamicDelta}&&\\
&\begin{array}{l}
\delta_{2\varsigma}(0,t)=0\\
\delta_{2\varsigma}(1,t)=\mathbfcal{L}_\mathbfcal{C}\left[\dot{U}(t)+\dot{\Psi}(t)\right]
\end{array} &&
\label{sect2:disDynamicBCs2}\\
&
\begin{array}{l}
\delta_1(\varsigma,0)=  \mathbfcal{L}_\mathbfcal{C} Q_{0}(\varsigma)\in \mathrm{H}^{4,N}(0,1) \\
\delta_2(\varsigma,0)=\theta \mathbfcal{L}_\mathbfcal{C} Q_{0\varsigma\varsigma}(\varsigma)\in \mathrm{H}^{2,N}(0,1)
\end{array} &&
\label{sect2:disDynamicBCs}
\end{align}

Presenting the second main result of this paper is preceded by the following instrumental lemma.

\vspace{.05cm}

\begin{mylem}\label{lemma2}
The functional
\begin{align}
V(\delta_1,\delta_2)=\theta a  \|\mathbfcal{L}\delta_1(1,t)\|_{{1}}+\frac{1}{2}\theta W_1  \|\mathbfcal{L}\delta_1(1,t)\|_{{2}}^2+\frac{1}{2}\int_0^1\delta_2(\xi,t)^T\mathbfcal{L}\delta_2(\xi,t)d\xi
\label{sect2:V}
\end{align}
being computed on the solutions ($\delta_1(\cdot,t),\delta_2(\cdot,t)$) of the BVP \eqref{sect2:disDynamicDelta}-\eqref{sect2:disDynamicBCs}, is equivalent to the $\mathrm{H}^{2,N}(0,1) \times \mathrm{H}^{0,N}(0,1)$ norm of these solutions in the sense that
\begin{flalign}
&\eta_1\left(
\|\delta_{1}(\cdot,t)\|_{\mathrm{H}^{2,N}}^2+
\|\delta_{2}(\cdot,t)\|_{\mathrm{H}^{0,N}}^2
\right)\leq V(\delta_1,\delta_2) \leq\eta_2 \left(
\|\delta_{1}(\cdot,t)\|_{\mathrm{H}^{2,N}}^2+
\|\delta_{2}(\cdot,t)\|_{\mathrm{H}^{0,N}}^2+\sum_{i=1}^N \|\delta_{1,i}(\cdot,t)\|_{\mathrm{H}^2}
\right)&&\label{sekjg}
\end{flalign}
 for an arbitrary solution ($\delta_1(\cdot,t),\delta_2(\cdot,t)$) of \eqref{sect2:disDynamicDelta}-\eqref{sect2:disDynamicBCs}, for all $t\geq0$, and for some positive constants $\eta_1$ and $\eta_2$.
\end{mylem}

\vspace{0.3cm}

\begin*{\bf Proof of Lemma \ref{lemma2}:} \label{prooflemma2}
It is preliminarily demonstrated that the condition
\begin{equation}
\alpha_1\cdot \tilde{V}(\delta_1,\delta_2)\leq
V(\delta_1,\delta_2)
\leq \alpha_2\cdot \tilde{V}(\delta_1,\delta_2)
\label{sect2:pf_lemma1}
\end{equation}
holds, where $\alpha_1$ and $\alpha_2$ are positive constants and $$\tilde{V}(\delta_1,\delta_2)=\theta a \|\delta_1(1,t)\|_{{1}}+\frac{1}{2}\theta W_1 \|\delta_1(1,t)\|_{{2}}^2+\frac{1}{2} \|\delta_2(\cdot,t)\|_{\mathrm{H}^{0,N}}^2.$$


By considering \eqref{sect1:laplacianProperty} and the second inequality of  \eqref{app1:prop1}, both specialized with ${x}={\delta_1}(1,t)$, one derives
\begin{equation}
\frac{\lambda_2}{\sqrt{N}}\cdot \|\delta_{1}(1,t)\|_{{1}}\leq \|\mathbfcal{L}\delta_{1}(1,t)\|_{{1}} \leq \|\mathbfcal{L}\|_{{1}}\|\delta_1(1,t)\|_{{1}}
\label{term1}
\end{equation}
Specializing \eqref{sect1:laplacianPropertySQUARE} with ${x}={\delta_1}(1,t)$, and \eqref{app1:QuadraticFormNorm} with ${x}={\delta_2}(\varsigma,t)$, one obtains
\begin{gather}
\lambda_2^2  \|\delta_{1}(1,t)\|_{{2}}^2\leq \|\mathbfcal{L}\delta_{1}(1,t)\|_{{2}}^2\leq
\lambda_N^2  \|\delta_{1}(1,t)\|_{{2}}^2
\label{term2}\\
\lambda_2\|\delta_2(\varsigma,t)\|_{2}^2\leq
 \delta_2(\varsigma,t)^T\mathbfcal{L}\delta_2(\varsigma,t)
\leq\lambda_N\|\delta_2(\varsigma,t)\|_{2}^2
\label{oiterm3}
\end{gather}

Noticing that, by construction, $\int_0^1 \|\delta_2(\xi,t)\|_{2}^2 d\xi = \|\delta_2(\cdot,t)\|_{\mathrm{H}^{0,N}}^2$,  the next estimate is derived after spatial integration of all terms in \eqref{oiterm3}
\begin{gather}
\lambda_2\|\delta_2(\cdot,t)\|_{\mathrm{H}^{0,N}}^2\leq
\int_0^1\delta_2(\xi,t)^T\mathbfcal{L}\delta_2(\xi,t)d\xi
\leq\lambda_N\|\delta_2(\cdot,t)\|_{\mathrm{H}^{0,N}}^2.
\label{term3}
\end{gather}

By \eqref{term1}-\eqref{term2} and \eqref{term3}, relation \eqref{sect2:pf_lemma1} is derived with the positive constants $\alpha_1 = \min \{\lambda_2/\sqrt{N},\lambda_2^2\} $ and $\alpha_2=\max \{\|\mathbfcal{L}\|_{{1}},\lambda_N^2, \lambda_N\}$. Furthermore, by \eqref{app1:prop1} and \eqref{app1:L2normVector}, functional $\tilde{V}(\delta_1,\delta_2)$ can be rewritten as follows:
\begin{align}
\tilde{V}(\delta_1,\delta_2)=&\sum_{i=1}^N \tilde{V}_i(\delta_{1i},\delta_{2i})
\label{vtildei}
\end{align}
where $$\tilde{V}_i=
\theta a \left|\delta_{1i}(1,t)\right|+\frac{1}{2}\theta W_1\delta_{1i}(1,t)^2+\frac{1}{2}\|\delta_{2i}(\cdot,t)\|_{\mathrm{H}^0}^2.$$ From that, by applying \cite[Lemma 2]{pisano2012boundary}, the next estimate
\begin{flalign}
&\beta_1\left( \|\delta_{1i}(\cdot,t)\|_{\mathrm{H}^2}^2+\|\delta_{2i}(\cdot,t)\|_{\mathrm{H}^0}^2\right)\leq \tilde{V}_i \leq\beta_2 \left( \|\delta_{1i}(\cdot,t)\|_{\mathrm{H}^2}^2+\|\delta_{2i}(\cdot,t)\|_{\mathrm{H}^0}^2+ \|\delta_{1i}(\cdot,t)\|_{\mathrm{H}^2}\right)&&\label{lmmkk}
\end{flalign}
holds for some positive constants $\beta_1$ and $\beta_2$. Then, by combining \eqref{vtildei} and \eqref{lmmkk}, and by applying definition \eqref{app1:L2normVector}, it results
\begin{flalign}
&\beta_1\left(\|\delta_{1}(\cdot,t)\|_{\mathrm{H}^{2,N}}^2+\|\delta_{2}(\cdot,t)\|_{\mathrm{H}^{0,N}}^2
\right)\leq \tilde{V}(\delta_1,\delta_2)  \leq\beta_2 \left( \|\delta_{1}(\cdot,t)\|_{\mathrm{H}^{2,N}}^2+ \|\delta_{2}(\cdot,t)\|_{\mathrm{H}^{0,N}}^2+\sum_{i=1}^N \|\delta_{1i}(\cdot,t)\|_{\mathrm{H}^2}
\right)&&\label{lmmkk2}
\end{flalign}

Finally, relation \eqref{sekjg}, being specified with the constants $\eta_1=\alpha_1\beta_1$ and $\eta_2=\alpha_2\beta_2$,  is straightforwardly derived by  combining \eqref{sect2:pf_lemma1}, \eqref{vtildei} and \eqref{lmmkk2}. Lemma~\ref{lemma2} is proved.
{\hfill $\square$}
\end*{}
\vspace{0.2cm}

Next theorem presents the second main result of this work.

\vspace{0.2cm}

\begin{mythm}\label{Theorem1}
Consider the perturbed multi-agent system  \eqref{sect1:barDynamicAVE}, \eqref{sect1:barICsAVE}, \eqref{sect1:barBCs}, with Assumptions \ref{assumption:1_distrurbance} and \ref{ass2}, communicating through an undirected connected static graph with Laplacian matrix $\mathbfcal{L}$. Let the boundary local interaction strategy \eqref{sect1:consensusProtocol0}-\eqref{sect1:consensusProtocol2} be applied, with the tuning parameters selected according to
\begin{equation}
a>b+\Pi,\quad b>\Pi,\quad W_1>0,\quad W_2>0,\quad W_3>0
\label{sect2:tuningRules}
\end{equation}
Then, condition \eqref{sect1:consensusDefinitionAVE} is achieved. {\hfill $\square$}
\end{mythm}

\vspace{0.2cm}

\begin*{\bf Proof of Theorem \ref{Theorem1}:}  See Appendix for the proof.

\end*{}

\vspace{0.15cm}

\begin{myrem}\label{remark:synchro}
{ In contrast to the average consensus result, outlined in Section II, in this case the steady-state common profile reached by the agents cannot be predicted \emph{a-priori} and it turns out to depend not only on the agents initial conditions but also on the actual controller parameters and disturbance vector. For this reason, the term $\lq\lq$robust synchronization" has been adopted in the present scenario to describe the underlying result, as opposed to the word "consensus" that mostly refers, in the literature, to situations where the steady state behaviour of the agents is determined a priori and the problem is that of enforcing it in a decentralized manner.}
{\hfill $\square$}
\end{myrem}

\section{Simulation Results}\label{sect5}

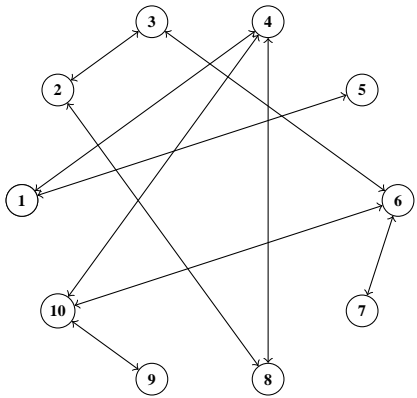
\begin{figure}[h]
\centering
\begin{tikzpicture}[scale=2.5,auto=center]
\tikzstyle{vertex}=[shape=circle,draw,fill=white,minimum size=12pt,inner sep=2pt,thin]
\foreach \name/\angle/\text in {P-1/180/1,
P-1/180/1,
P-2/144/2,
P-3/108/3,
P-4/72/4,
P-5/36/5,
P-6/0/6,
P-7/-36/7,
P-8/-72/8,
P-9/-108/9,
P-10/-144/10
}
\node[vertex] (\name) at (\angle:1cm) {\tiny$\bf\text$};
\foreach \from/\to in {1/4,1/5,2/3,3/6,4/10,2/8,6/7,9/10,4/8,6/10}
{ \draw [<->,thin] (P-\from) -- (P-\to);}
\end{tikzpicture}
\caption{\label{image:RETE} The considered network topology.}
\end{figure}

In the present section, simulation results are presented to illustrate the performance of the proposed protocols. The connected network of $N=10$ agents displayed in Figure~\ref{image:RETE} is considered, with the diffusivity parameter $\theta=1$.

For solving the resulting closed-loop system of coupled PDEs, the spatial domain $\varsigma \in [0,1]$ has been discretized by the standard finite-difference approximation method
considering $n=30$ uniformly spaced solution nodes. 
The resulting finite dimensional system of coupled ODEs is then solved by means of the Euler fixed-step solver with sampling-step $T_s= 10^{-4}$.
%

\subsection{Average consensus} \label{sect5.1}

\begin{figure}[!t]
\centering
\includegraphics[width=7cm]{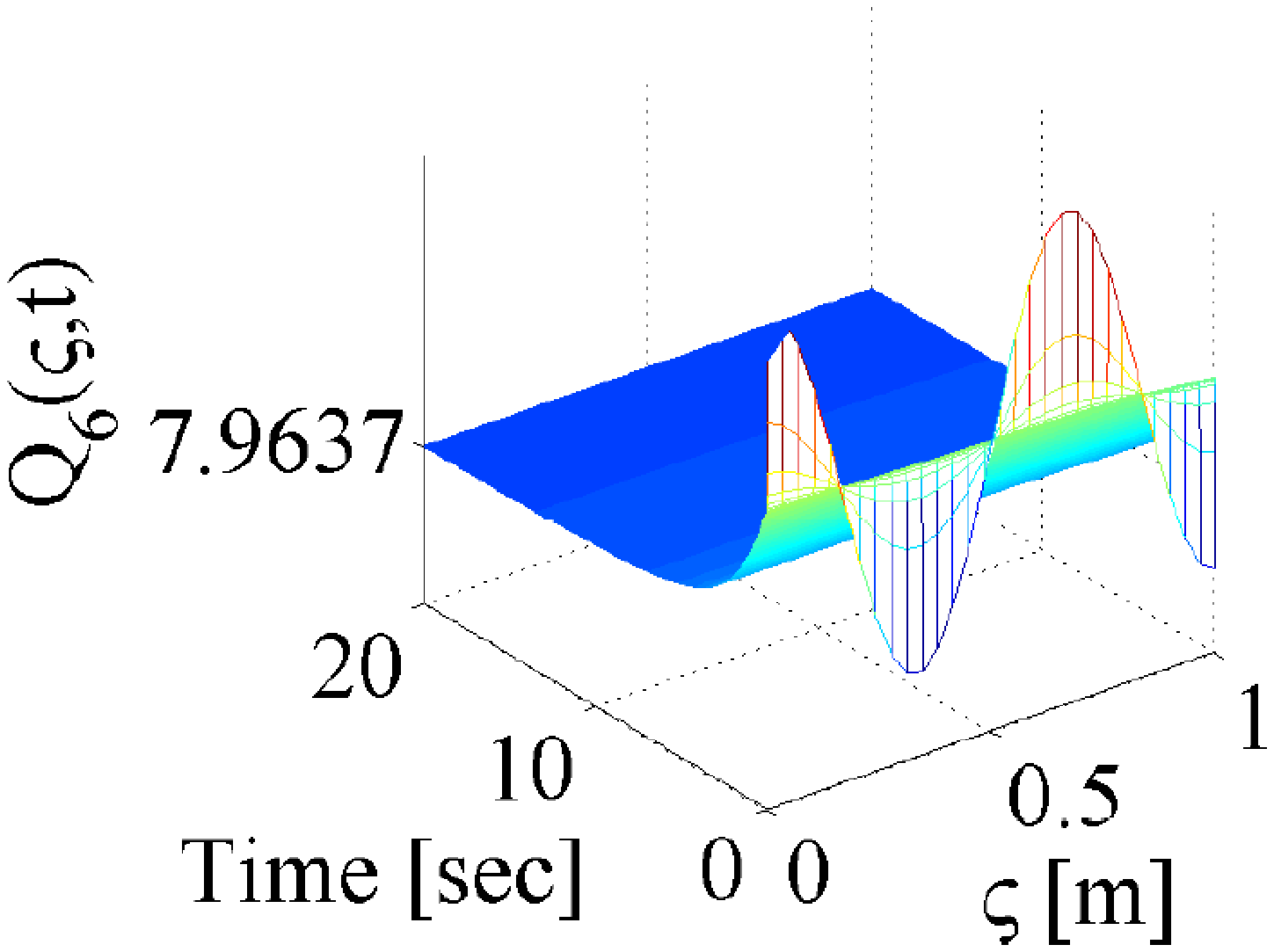}
\includegraphics[width=7cm]{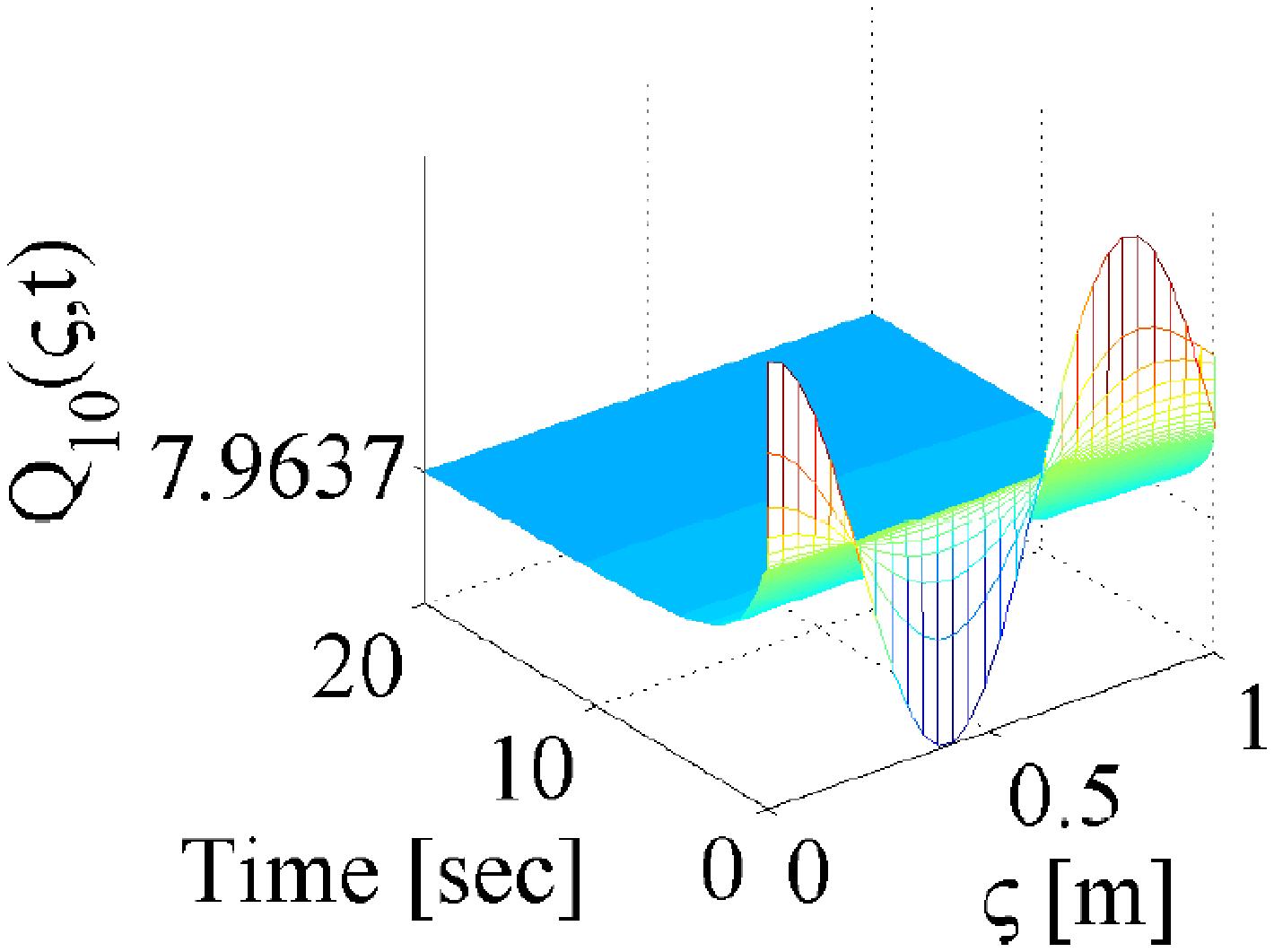}
\caption{TEST~1: Spatiotemporal profiles $Q_6(\varsigma,t)$ (left) and $Q_{10}(\varsigma,t)$ (right).\label{image:agent5nonconstave}}
\end{figure}

\begin{figure}[h]
\centering
\includegraphics[width=8cm]{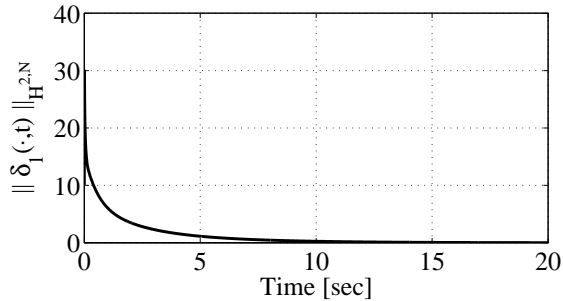}
\caption{TEST~1: Time evolution of the
disagreement vector norm $\|\delta_{1}(\cdot,t)\|_{\mathrm{H}^{2,N}}$.\label{image:V_tave}}
\end{figure}

System \eqref{sect1:barDynamicAVE}-\eqref{sect1:barBCsAVE}, coupled with the local interaction protocol \eqref{sect1:consensusProtocol0AVE}, is under investigation.
In the first simulation run (TEST~1), spatially varying ICs have been selected as follows: $Q_1(\varsigma,0)=10+\omega_1\cos(3\pi\varsigma)$, $Q_2(\varsigma,0)=10+\omega_2\cos(3\pi\varsigma)$,
$Q_3(\varsigma,0)=8+\omega_3\cos(3\pi\varsigma)$,
$Q_4(\varsigma,0)=10+\omega_4\cos(3\pi\varsigma)$,
$Q_5(\varsigma,0)=6+\omega_5\cos(3\pi\varsigma)$,
$Q_6(\varsigma,0)=10+\omega_6\cos(3\pi\varsigma)$,
$Q_7(\varsigma,0)=10+\omega_7\cos(3\pi\varsigma)$,
$Q_8(\varsigma,0)=-5+\omega_8\cos(3\pi\varsigma)$, $Q^9(\varsigma,0)=10+\omega_9\cos(3\pi\varsigma)$,
$Q_{10}(\varsigma,0)=10+\omega_{10}\cos(2.5\pi\varsigma)$, with $\omega_i=1+4(i-1)/9$, $i=1,2\dots,10$. The corresponding spatial average $Q^*$, evaluated according to \eqref{sect1:consensusDefinitionAVE3}, is $Q^*=7.9637$. Figure~\ref{image:agent5nonconstave} shows the spatiotemporal evolutions of the states $Q_6(\varsigma,t)$ and $Q_{10}(\varsigma,t)$. The steady-state profile of both agents is constant and takes the expected pre-computed value $Q^*$. Figure~\ref{image:V_tave} displays the time evolution of the disagreement vector norm $\|\delta_{1}(\cdot,t)\|_{\mathrm{H}^{2,10}}$, which tends to zero thereby confirming that all agents reach a common steady-state profile.

\subsection{Robust Synchronization}\label{sect5.2}

The performance of the robust synchronization protocol \eqref{sect1:consensusProtocol0}-\eqref{sect1:consensusProtocol2}, presented in Section~\ref{sect4}, is now verified with reference to the perturbed PDEs \eqref{sect1:barDynamicAVE}, \eqref{sect1:barICsAVE}, \eqref{sect1:barBCs}. The entries $\psi_i(t)$ of the disturbance vector $\Psi(t)$ are selected as { $\psi_i(t)=4 k_i\cdot t+\sin\left(k_i\pi t\right)$ ($i=1,2,...,10$)}, where the coefficients $k_i$ are randomly chosen in the interval [0,2]. The considered disturbance, { which is unbounded in magnitude as time grows}, meets the restriction \eqref{sect1:disturbanceCondition} with a constant upper-bound constant $\Pi = 2\pi+8$. The chosen ICs are
\begin{equation}
Q_i(\varsigma,0)=10+(i-4.5)\cos(4\pi\varsigma)
\quad \quad i =1,2,...,10.\nonumber
\end{equation}

\begin{figure}[!t]
\centering
\includegraphics[width=7cm]{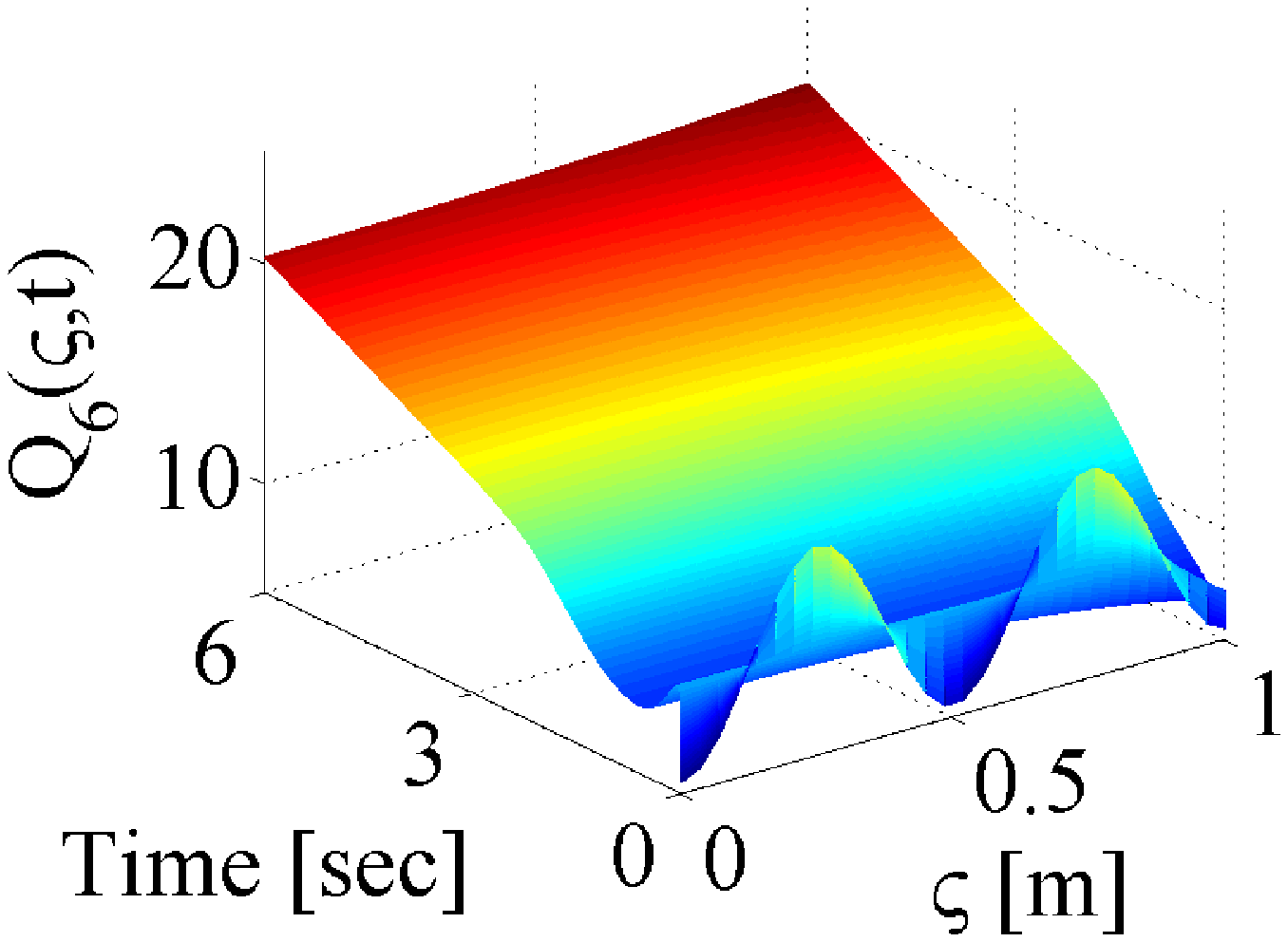}
\includegraphics[width=7cm]{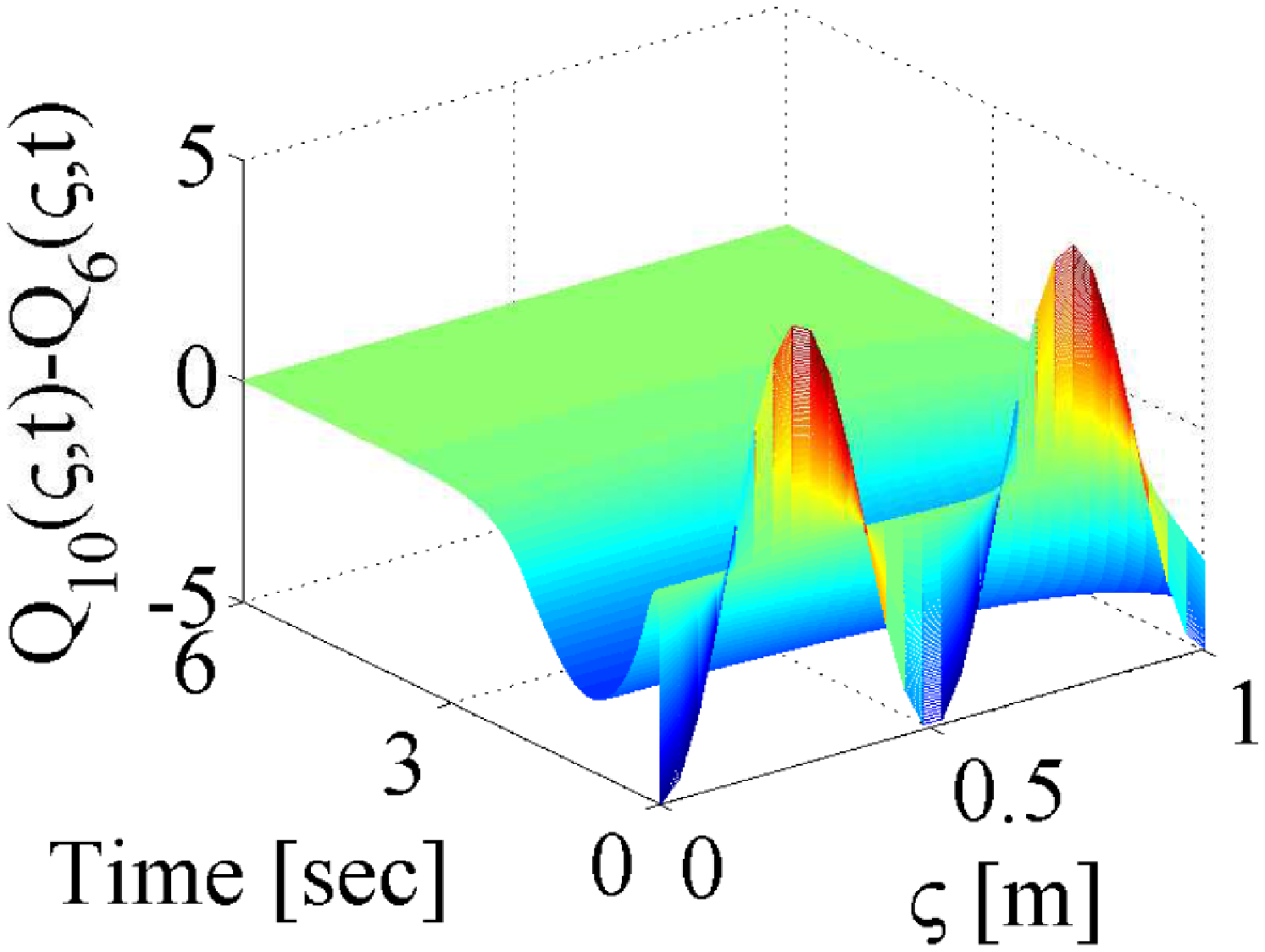}
\caption{TEST~2: Spatiotemporal profiles $Q_6(\varsigma,t)$ (left) and $Q_{6}(\varsigma,t)-Q_{10}(\varsigma,t)$ (right).\label{image:agent5}}
\centering
\includegraphics[width=7cm]{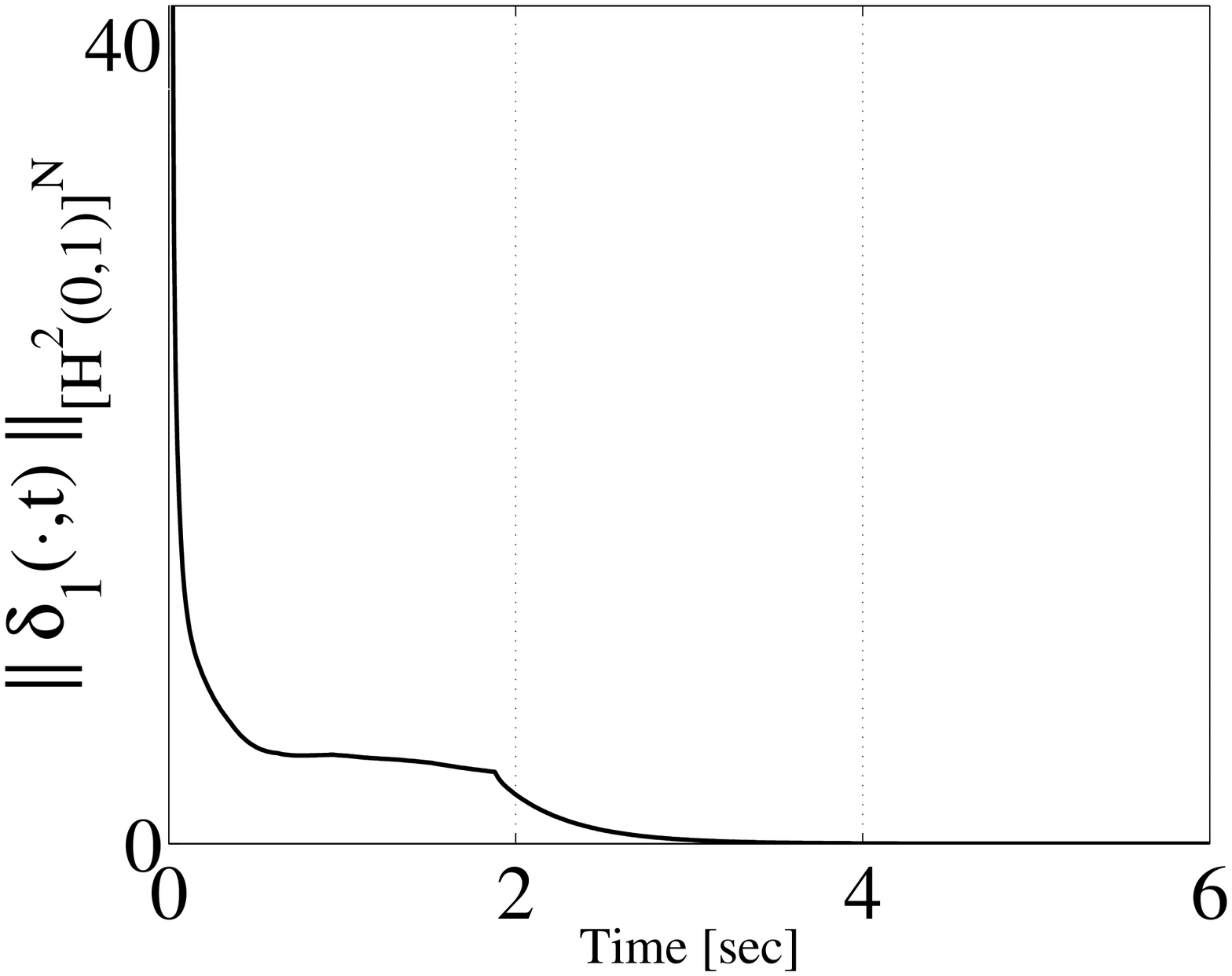}\includegraphics[width=7cm]{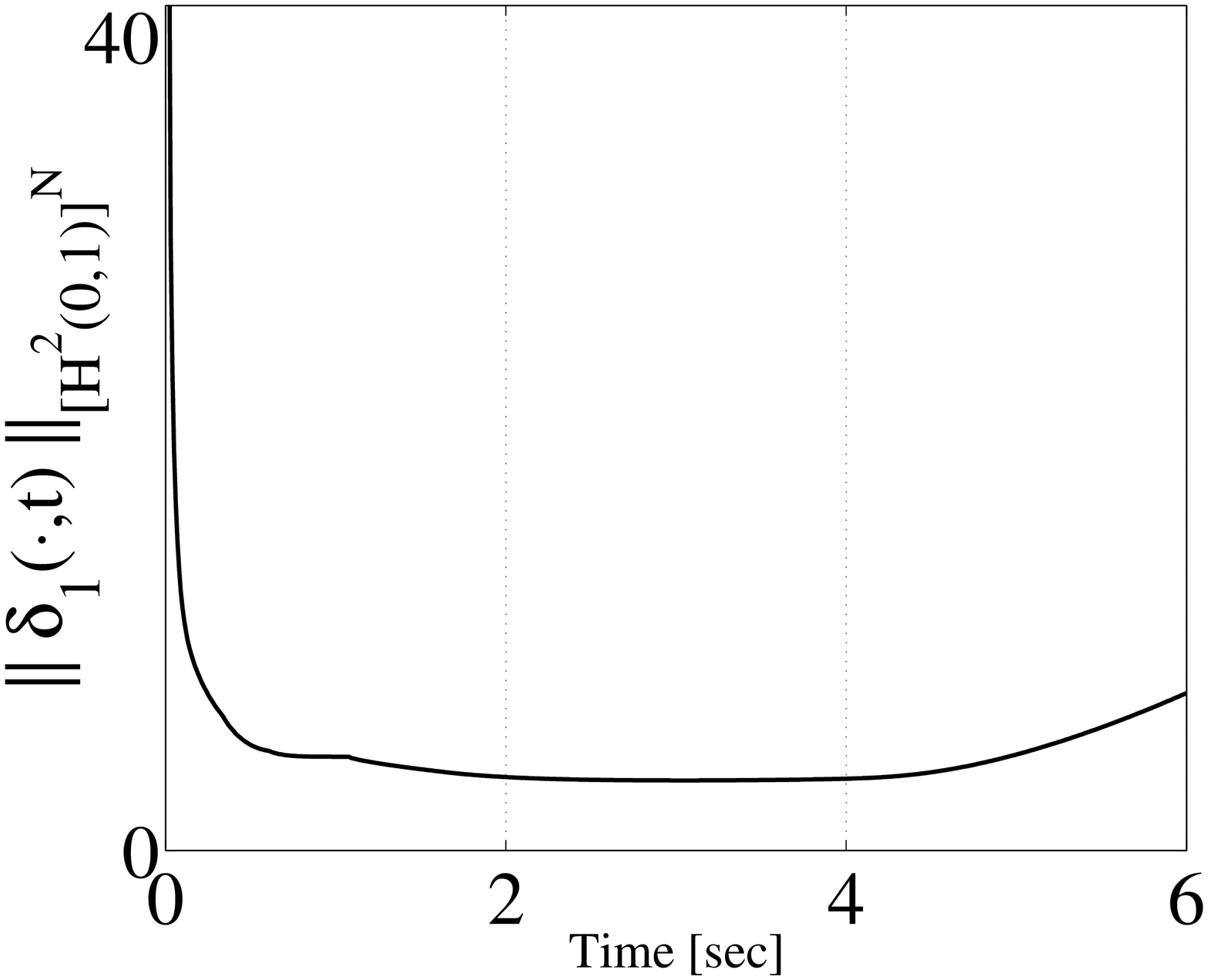}
\caption{Time evolution of the
norm $\|\delta_{1}(\cdot,t)\|_{\mathrm{H}^{2,N}}$ in TEST 2 (left) and TEST 3 (right).\label{image:V_t}}
\end{figure}

The tuning parameters $a=40, b=20, W_1=W_2=W_3=5$, were chosen according to Theorem~\ref{Theorem1}. This simulation run is referred to as TEST 2. Figure~\ref{image:agent5} shows the spatiotemporal evolution of the state $Q_6(\varsigma,t)$ and of the state mismatch $Q_{10}(\varsigma,t)-Q_6(\varsigma,t)$. It is clear that both states converge towards the same steady-state profile. The time evolution of the disagreement vector norm $\|\delta_{1}(\cdot,t)\|_{\mathrm{H}^{2,10}}$, shown in Figure \ref{image:V_t}-left, tends to zero as shown in the Theorem~\ref{Theorem1}. { To verify the conservativeness of the approach, another simulation run, called TEST 3, was made where the terms $\alpha_i t^2$, $i=1,2,\ldots,10$  ($\alpha_i$ being randomly chosen constants in the interval $[0 , 20]$) has been added to the  disturbance entries $\psi_i(t)$ used in TEST 2. Due to the insertion of these additional terms, not only the magnitudes of the disturbances are increasing with time but also those of their derivatives are. Thus, the tuning conditions \eqref{sect2:tuningRules} are deliberately violated. Figure\ref{image:V_t}-right depicts the resulting divergence trend of the  disagreement vector norm $\|\delta_{1}(\cdot,t)\|_{\mathrm{H}^{2,10}}$, thereby supporting the theoretical results.}

\section{Conclusion}\label{sect6}


In this work, an infinite-dimensional counterpart of the well-known finite-dimensional average consensus algorithm has been derived with reference to an unperturbed network of diffusion processes under a linear decentralized local interaction policy. Along with this, the problem of guaranteeing the asymptotic agreement among the agents' states while rejecting a class of persistent and possibly unbounded disturbances has been tackled by devising a nonlinear local interaction policy based on the second-order sliding-mode control approach.

Future activities will be targeted to relaxing the topological restrictions on the network structure by covering, e.g., directed and possibly switching communication graphs. Further investigation is called for the extension of these results to more general classes of (possibly non identical) distributed parameters agents' dynamics and additionally to more general consensus-based problems in the infinite dimensional setting such as e.g. leader following.

%


\ifCLASSOPTIONcaptionsoff
  \newpage
\fi


%

\bibliographystyle{IEEEtran}
\bibliography{autosam}

\IEEEtriggercmd{\enlargethispage{-5in}}

\appendix
\begin{center}
\textbf{Proof of Theorem \ref{Theorem1}}
\end{center}

%
%
%


Consider the Lyapunov function \eqref{sect2:V}. Its time derivative is
\begin{flalign}
\dot{V}(t)=&\theta a\delta_2(1,t)^T\mathbfcal{L} \mathrm{Sign}(\mathbfcal{L}\delta_1(1,t))+\theta W_1 \delta_2(1,t)^T\mathbfcal{L}^2\delta_1(1,t)+\int_0^1 \delta_2(\xi,t)^T\mathbfcal{L}\delta_{2t}(\xi,t)d\xi
\label{sect2:Vdot1}
\end{flalign}

Substituting \eqref{sect2:disDynamicDelta} into the last term of \eqref{sect2:Vdot1}, and performing integration by parts in light of \eqref{sect2:disDynamicBCs2}, yield
\begin{flalign}
&\int_0^1 \delta_2(\xi,t)^T\mathbfcal{L}\delta_{2t}(\xi,t)d\xi=\theta\int_0^1 \delta_2(\xi,t)^T\mathbfcal{L}\delta_{2\xi\xi}(\xi,t)d\xi \nonumber \\
& = \theta \delta_2(1,t)^T\mathbfcal{L}\delta_{2\varsigma}(1,t)-\theta\int_0^1\delta_{2\xi}(\xi,t)^T\mathbfcal{L}\delta_{2\xi}(\xi,t)d\xi
\label{sect2:term1AUU}
\end{flalign}
Owing on the spatial differentiation of equation \eqref{sect2:1tlamn1AVE}, relation \eqref{app1:QuadraticFormNorm} specified with ${x}={\delta_{2\varsigma}}$ holds true.  Thus, the last integral term of \eqref{sect2:term1AUU}  can be estimated as follows
\begin{equation}
-\theta \int_0^1 \delta_{2\xi}(\xi,t)^T\mathbfcal{L} \delta_{2\xi}(\xi,t) d\xi  \leq -\theta\lambda_2\|\delta_{2\varsigma}(\cdot,t)\|_{\mathrm{H}^{0,N}}^2
\label{sect2:term1}
\end{equation}

Substituting the  BCs \eqref{sect2:disDynamicBCs2} and the controller equations \eqref{sect1:consensusProtocol0}-\eqref{sect1:consensusProtocol2} into the right-hand side of \eqref{sect2:term1AUU} one derives
\begin{flalign}
& \theta \delta_2(1,t)^T\mathbfcal{L}\delta_{2\varsigma}(1,t)=-\theta a\cdot \delta_2(1,t)^T\mathbfcal{L} \mathrm{Sign}\left(\mathbfcal{L}\delta_1(1,t)\right) -\theta b  \|\mathbfcal{L}\delta_2(1,t)\|_{{1}} -\theta W_1 \delta_2(1,t)^T\mathbfcal{L}^2\delta_1(1,t) && \nonumber \\
&-\theta W_2  \|\mathbfcal{L}\delta_2(1,t)\|_{{2}}^2 -\theta W_3 \delta_2(1,t)^T\mathbfcal{L}Q_t(1,t)+\theta \delta_2(1,t)^T\mathbfcal{L}\dot{\Psi}(t)
\label{sect2:term2}
\end{flalign}

Employing \eqref{sect2:LLCDIAt}, evaluated at $\varsigma=1$, and considering \eqref{app1:QuadraticFormNorm} specified with $x=\delta_2(1,t)$, one gets
\begin{equation}
- \theta W_3 \delta_2(1,t)^T\mathbfcal{L}Q_t(1,t)\leq -\theta W_3 \lambda_2\cdot\|\delta_2(1,t)\|_{{2}}^2 \label{sect2:lb1}
\end{equation}

By applying \eqref{app1:HolderInequality}, specified with $x=\dot{\Psi}(t)$, $y=\mathbfcal{L}  \delta_2(1,t)$, $p=\infty$, $q=1$, and taking into account \eqref{sect1:disturbanceCondition}, the magnitude of the last term in the right hand side of \eqref{sect2:term2} is estimated as follows
\begin{equation}
|\theta \delta_2(1,t)^T\mathbfcal{L}\dot{\Psi}(t)| = \theta |\dot{\Psi}^T(t) \mathbfcal{L}  \delta_2(1,t)| \leq \theta \Pi \|  \mathbfcal{L}  \delta_2(1,t)\|_1 \label{sect2:lb2}
\end{equation}

Substituting \eqref{sect2:term1AUU}-\eqref{sect2:lb2} into \eqref{sect2:Vdot1}, and considering relation $\lambda_2^2  \|\delta_{1}(1,t)\|_{{2}}^2\leq \|\mathbfcal{L}\delta_{1}(1,t)\|_{{2}}^2$,
which  is verified by applying \eqref{sect1:laplacianPropertySQUARE} specialized with ${x}={\delta_2}(1,t)$,   the next inequality
\begin{flalign}
\dot{V}(t)\leq&-\theta\cdot(b-\Pi)\cdot\|\mathbfcal{L}\delta_2(1,t)\|_{{1}}-\theta\lambda_2\cdot\|\delta_{2\varsigma}(\cdot,t)\|_{\mathrm{H}^{0,N}}^2\nonumber\\
&-\theta W_2\lambda_2^2 \cdot\|\delta_2(1,t)\|_{{2}}^2-\theta W_3 \lambda_2\cdot\|\delta_2(1,t)\|_{{2}}^2
\label{sect2:VdotUpperEstimation}
\end{flalign}
is concluded after some straightforward manipulations. Due to \eqref{sect2:VdotUpperEstimation} and \eqref{sect2:tuningRules}, the time derivative of the Lyapunov functional $ V(t)$,  being computed along the solutions of the closed-loop system, is negative semi-definite, and $V(t)$ is thus a non-increasing function of time. Thereby, the set
\begin{flalign}
\mathcal{D}_{R}^{V}=&\left\lbrace \left( \delta_1,\delta_2\right)\in \mathrm{H}^{2,N}\times \mathrm{H}^{0,N}: V\left(\delta_1,\delta_2\right)\leq R\right\rbrace
\label{sect2:CompactSet}
\end{flalign}
specified for an arbitrary  $R \geq V(0)$, is invariant.  
By exploiting the invariance of the domain  $ \mathcal{D}^{V}_R $, the next estimates are derived by straightforward manipulations of the inequality $V(\cdot) \leq R$ in light of \eqref{sect2:V}, \eqref{term2} and \eqref{term3}:
\begin{eqnarray}
\|\mathbfcal{L}\bm{\delta}_1(1,t)\|_{{1}}&\leq& R/\theta a
\label{sect2:bounds1}\\
\|\bm{\delta}_1(1,t)\|_{{2}}^2&\leq& 2R/\theta W_1\lambda^2_2 \label{sect2:bounds2}\\
\|\bm{\delta}_2(\cdot,t)\|_{\mathrm{H}^{0,N}}^2&\leq& 2 R/\lambda_2
\label{sect2:bounds3}
\end{eqnarray}

%

Now consider the $\lq\lq$augmented" functional
\begin{flalign}
V_R(t)&=V(t)+\kappa_R\cdot \bar{V}(t) \label{sect2:VR} \\
\bar{V}(t)&=\frac{1}{2}\theta W_2\cdot\|\mathbfcal{L}\delta_1(1,t)\|_{{2}}^2+\int_0^1\delta_1(1,t)^T\mathbfcal{L}\delta_2(\xi,t)d\xi\label{sect2:U}
\end{flalign}
where $\kappa_R$ is a sufficiently small positive constant to subsequently be specified. The next estimation holds
\begin{equation}
\frac{1}{2}\theta W_2\cdot\|\mathbfcal{L}\delta_1(1,t)\|_{{2}}^2\geq \frac{1}{2}\theta W_2\lambda_2^2 \cdot \|\delta_1(1,t)\|_{{2}}^2,
\label{sect2:Ubound2}
\end{equation}
whereas by \eqref{app1:prop1}, \eqref{app1:HolderInequality} and \eqref{sect2:bounds1}, the second term in the right-hand side of \eqref{sect2:U} is manipulated as
\begin{equation}
\int_0^1\delta_1(1,t)^T\mathbfcal{L}\delta_2(\xi,t)d\xi \geq-\frac{1}{2}\left(\frac{R}{\theta a}\cdot
\|\mathbfcal{L}\delta_1(1,t)\|_{{1}}+\|\delta_2(\cdot,t)\|_{\mathrm{H}^{0,N}}^2
\right).
\label{sect2:Ubound1}
\end{equation}
Considering \eqref{sect2:Ubound2} and \eqref{sect2:Ubound1} along with \eqref{sect2:VR}, \eqref{sect2:V}, yield
\begin{equation}
V_R(t)\geq\left[\theta a -\frac{\kappa_R R}{2\theta a}\right]   \|\mathbfcal{L}\delta_1(1,t)\|_{{1}}+\frac{\lambda_2-\kappa_R}{2} \|\delta_2(\cdot,t)\|_{\mathrm{H}^{0,N}}^2+\frac{1}{2}\theta\lambda_2^2\left(W_1+\kappa_R W_2\right)  \|\delta_1(1,t)\|_{{2}}^2.
\label{sect2:Ulowerbound}
\end{equation}

Thus, the positive definitiveness of $V_R(t)$ is guaranteed by selecting the positive constant $\kappa_R$ small enough according to $$\kappa_R\leq\min
\begin{Bmatrix}
\frac{2 \theta^2 a^2}{R},\lambda_2
\end{Bmatrix}$$.
In particular, in the invariant domain  $ \mathcal{D}^{V}_R $  the augmented functional $V_R(t)$ turns out to be lower estimated in terms of $V(t)$:
\begin{equation}
V_R(t)\geq\min \left\{\frac{\theta a}{\theta a - \frac{\kappa_R R}{2 \theta a}},\frac{\lambda_2}{\lambda_2-\kappa_R},\frac{W_1}{W_1+\kappa_RW_2}\right\} V(t).
\label{sect2:vrlowerest}
\end{equation}

Differentiating \eqref{sect2:VR} along the solutions of  \eqref{sect2:disDynamicDelta}-\eqref{sect2:disDynamicBCs}, and exploiting the identity $\mathbfcal{L}  \mathbfcal{L}_{\mathbfcal{C}}=\mathbfcal{L}$ in \eqref{LcLAVE}, it yields
\begin{flalign}
\dot{V}_R(t)&=\dot{V}(t)+\kappa_R \theta W_2\delta_2(1,t)\mathbfcal{L}^2\delta_1(1,t)+\kappa_R \int_0^1 \theta\delta_1(1,t)^T\mathbfcal{L} \delta_{2,\xi\xi}(\xi,t)d\xi+\kappa_R \int_0^1 \delta_{2}(1,t)^T\mathbfcal{L} \delta_2(\xi,t)d\xi.
\label{sect2:Vrdot}
\end{flalign}
Employing \eqref{sect1:consensusProtocol0}-\eqref{sect1:consensusProtocol2} and the BCs \eqref{sect2:disDynamicBCs2} one gets
\begin{flalign}
&\int_0^1 \theta\delta_1(1,t)^T\mathbfcal{L}\delta_{2,\xi\xi}(\xi,t)d\xi=
\theta\delta_1(1,t)^T\mathbfcal{L}\delta_{2,\varsigma}(1,t) =-\theta a\cdot \|\mathbfcal{L}\delta_1(1,t)\|_{{1}}-\theta b\cdot \delta_1(1,t)\mathbfcal{L}\mathrm{Sign}(\mathbfcal{L}\delta_2(1,t))\nonumber\\
&-\theta W_1\cdot \|\mathbfcal{L}\delta_1(1,t)\|_{{2}}^2-\theta W_2\cdot \delta_1(1,t)^T\mathbfcal{L}^2\delta_2(1,t)-\theta W_3\cdot \delta_1(1,t)^T\mathbfcal{L}\delta_2(1,t)+\theta\cdot \delta_1(1,t)^T\mathbfcal{L}\dot{\Psi}(t).&&
\label{sect2:Udot_I1}
\end{flalign}
Then, utilizing \eqref{app1:HolderInequality} and \eqref{sect2:bounds2}, the  sign-indefinite terms in \eqref{sect2:Udot_I1} are estimated as follows
\begin{align}
&\mid-\delta_1(1,t)\mathbfcal{L}\mathrm{Sign}(\mathbfcal{L}\delta_2(1,t))\mid\leq \|\mathbfcal{L}\delta_{1}(1,t)\|_{1}\label{est1}~,&&\\
&|-\delta_1(1,t)^T\mathbfcal{L}\delta_2(1,t)|\leq
\|\delta_1(1,t)\|_{{2}}\|\mathbfcal{L}\delta_2(1,t)\|_{{2}} \nonumber\\
&\leq \|\delta_1(1,t)\|_{{2}} \|\mathbfcal{L}\delta_2(1,t)\|_{{1}}
\leq {\small\sqrt{\frac{2R}{\theta W_1\lambda_2^2}}}\cdot \|\mathbfcal{L}\delta_2(1,t)\|_{{1}}~,
\label{sect2:upperEstimationLastTermVdotR}\\
&\mid \delta_1(1,t)^T\mathbfcal{L}\dot{\Psi}(t) \mid \leq \|\Psi(t)\|_{\infty}\|\mathbfcal{L}\delta_{1}(1,t)\|_{1}\label{term4}~.&&
\end{align}

By \eqref{sect2:bounds3} and the Holder integral inequality, the last integral term in the right hand side of \eqref{sect2:Vrdot} is estimated as
\begin{flalign}
&\left|\int_0^1 \delta_2(1,t)^T\mathbfcal{L}\delta_2(\xi,t)d\xi\right|\leq \sqrt{\frac{2R}{\lambda_2}}\cdot\|\mathbfcal{L}\delta_{2}(1,t)\|_{{1}} \label{sect2:80z}
\end{flalign}

By \eqref{sect2:Vrdot}-\eqref{sect2:80z}, and considering \eqref{sect1:laplacianPropertySQUARE} and \eqref{sect2:VdotUpperEstimation}, one manipulates \eqref{sect2:Vrdot} as follows:
\begin{small}
\begin{flalign}
&\dot{V}_R(t)\leq-\theta\left(b-\Pi-\kappa_R\sqrt{2R/\theta^2\lambda_2} \right)
\|\mathbfcal{L}\delta_2(1,t)\|_{{1}}+\kappa_R W_3\theta\sqrt{{2R/\theta W_1\lambda^2_2}}\cdot\|\mathbfcal{L}\delta_2(1,t)\|_{{1}}\nonumber\\
&-\kappa_R\theta\left(a-b-\Pi\right)\cdot\|\mathbfcal{L}\delta_1(1,t)\|_{{1}}-\theta\lambda_2\cdot\|\delta_{2,\varsigma}(\cdot,t)\|_{\mathrm{H}^{0,N}}^2 -\theta (W_2\lambda_2^2+W_3\lambda_2)\cdot\|\delta_2(1,t)\|_{{2}}^2-\kappa_R\theta W_1\lambda_2^2\cdot\|\delta_1(1,t)\|_{{2}}^2\label{sect2:Vrdot_upper_step2}
\end{flalign}
\end{small}

By Lemma~\ref{lemma1:pisanoResult}, specialized for ${b}(\cdot)=\delta_{2}(\cdot)$ and $i=1$, 
the next estimate
 \begin{flalign}
-\theta(W_2\lambda_2^2+W_3\lambda_2) \|\delta_2(1,t)\|_{{2}}^2-\theta\lambda_2\|\delta_{2\varsigma}(\cdot,t)\|_{{2}}^2 \leq-c_4\|\delta_2(\cdot,t)\|_{ \mathrm{H}^{0,N}}^2
\label{sect2:lastTrick}
\end{flalign}
 is obtained with $$c_4=\theta \lambda_2 \min\left\lbrace 1, (W_2\lambda_2+W_3)\right\rbrace $$. By substituting \eqref{sect2:lastTrick} into the right hand side of \eqref{sect2:Vrdot_upper_step2}, the inequality
\begin{flalign}
\dot{V}_R(t)\leq&-c_1\cdot \|\mathbfcal{L}\delta_1(1,t)\|_{{1}}-c_2\cdot
\|\mathbfcal{L}\delta_2(1,t)\|_{{1}} -c_3\cdot\|\delta_1(1,t)\|_{{2}}^2-c_4\cdot\|\delta_2(\cdot,t)\|_{ \mathrm{H}^{0,N}}^2
\label{sect2:VRdot_Final}
\end{flalign}
is finally obtained, with the coefficients $$c_1=\kappa_R\theta\left(a-b-\Pi\right), \;\;\; c_3=\kappa_R\theta W_1\lambda_2^2,$$, $$c_2={\small\theta\left(b-\Pi-\kappa_R\left(\sqrt{\frac{2R}{\theta^2\lambda_2}}+\frac{W_3}{\lambda_2}\sqrt{\frac{2R}{\theta W_1}}\right)\right)}.$$ It is clear that due to the proposed specifications of constants $c_1, c_2, c_3$, all terms, appearing in the right-hand side of \eqref{sect2:VRdot_Final},
are nonpositive provided that the tuning conditions \eqref{sect2:tuningRules}, imposed
on the controller parameters, hold and the next more restrictive condition $$\kappa_R\leq\min
\begin{Bmatrix}
\frac{2 \theta^2 a^2}{R},\lambda_2,\frac{b-\Pi}{\sqrt{\frac{2R}{\theta^2\lambda_2}}+\frac{W_3}{\lambda_2}\sqrt{\frac{2R}{\theta W_1}}}
\end{Bmatrix}$$ is additionally satisfied. It then follows from \eqref{sect2:VRdot_Final} that
\begin{equation}
\dot{V}_R(t)\leq-\gamma_1 ( \|\mathbfcal{L}\delta_1(1,t)\|_{{1}}
+\|\delta_1(1,t)\|_{{2}}^2 +\|\delta_2(\cdot,t)\|_{\mathrm{H}^{0,N}}^2 )
\label{sect2:VRdot_Final23}
\end{equation}
with $\gamma_1=\min\lbrace c_1, c_3, c_4 \rbrace > 0$. By \eqref{sect2:V} combined with the first inequalities of \eqref{app1:QuadraticFormNorm} and \eqref{sect1:laplacianPropertySQUARE} it yields
\begin{equation}
V(t)\leq\theta a \|\mathbfcal{L}\delta_1(1,t)\|_1+\frac{1}{2}\theta W_1 \lambda_N^2\|\delta_{1}(1,t)\|_{2}^2+\frac{1}{2}\|
\delta_{2}(\cdot,t)
\|_{\mathrm{H}^{0,N}},
\nonumber
\end{equation}
whereas by \eqref{app1:QuadraticFormNorm}, \eqref{sect2:bounds1} and \eqref{app1:prop1}, along with property \eqref{app1:HolderInequality} specialized with $x=\mathbfcal{L}\delta_1(1,t)$ and $y=\delta_{2}(\varsigma, t)$, one derives that
\begin{equation}
\bar{V}(t)\leq\frac{1}{2}\theta W_2\lambda_{N}^2\|\delta_{1}(1,t)\|_{2}^2+\frac{R}{2 \theta a}\|\mathbfcal{L}\delta_1(1,t)\|_{1}+\frac{1}{2}\|\delta_2(\cdot,t)\|_{\mathrm{H}^{0,N}}
\nonumber
\end{equation}
Finally, substituting the last two estimations in \eqref{sect2:VR} one obtains
\begin{equation}
V_{R}(t)\leq\gamma_2( \|\mathbfcal{L}\delta_1(1,t)\|_{{1}}
+\|\delta_1(1,t)\|_{{2}}^2 +\|\delta_2(\cdot,t)\|_{[\mathrm{H}^{0}(0,1)]^n}^2 )
\label{sect2:VR_Final23}
\end{equation}
where $$\gamma_2=\min\lbrace \theta a-\frac{\kappa_R R}{2\theta a}, \frac{\left(\lambda_N-\kappa_R\right)}{2}, \frac{\theta \lambda_N^2\left(W_1+\kappa_R W_2\right)}{2} \rbrace > 0.$$  Thus, one derives from \eqref{sect2:VRdot_Final23} and \eqref{sect2:VR_Final23} that $$\dot{V}_R(t)\leq -\rho_R\cdot V_R(t),  \;\;\;\; \rho_R=\gamma_1/\gamma_2,$$ thereby concluding the exponential decay of $V_R(t)$, initialized within the invariant set $\mathcal{D}_R^V$ in \eqref{sect2:CompactSet}.

To complete the proof, it remains to note that due to the upper estimate \eqref{sect2:vrlowerest} 
the functional ${V}(t)$ decays, too. By applying Lemma~\ref{lemma2}, the local asymptotic stability of \eqref{sect2:disDynamicDelta}-\eqref{sect2:disDynamicBCs} is then established in the space $H^{2,N}(0,1) \times H^{0,N}(0,1) $ for the initial set \eqref{sect2:CompactSet}. Since
\eqref{sect2:CompactSet} can be specified with an arbitrarily large $R>0$, thus capturing an arbitrarily large initial domain, and the tuning conditions \eqref{sect2:tuningRules} do not depend on $R$, the
{global} asymptotic stability is then concluded in the space $\mathrm{H}^{2,N}(0,1) \times \mathrm{H}^{0,N}(0,1) $. It follows from \eqref{sekjg} that $\|\delta_1(\cdot,t)\|_{\mathrm{H}^{2,N}}$ asymptotically vanishes too, which results in the  following component-wise relations
\begin{align}
& \lim_{t \rightarrow \infty} \|{\delta}_{1i}(\cdot,t)\|_{\mathrm{H}^2} = 0, &\forall ~ i\in \mathbfcal{V},\label{delta1convz}&  \qquad\qquad
\end{align}

{ It is well known \cite{henry} that the Sobolev space $H^2(0,1)$ is continuously embedded in the Banach space $C(0,1)$ equipped with the supremum norm. In other words, there exists a constant $M>0$ such that
\begin{align}
& sup_{\xi \in [0,1]} |{\delta}_{1i}(\xi,t)| \leq M \|{\delta}_{1i}(\cdot,t)\|_{\mathrm{H}^2}, &\forall ~ i\in \mathbfcal{V},\label{embed}
\end{align}

Thus, one concludes the spatially point-wise
decay of all entries of $\bm{\delta}_{1}(\cdot,t)$. This property, coupled to the identity
$ Q_i(\varsigma,t)- Q_j(\varsigma,t) = {\delta}_{1i}(\varsigma,t)-{\delta}_{1j}(\varsigma,t)$, yields \eqref{sect1:consensusDefinitionAVE}}. The proof of Theorem~\ref{Theorem1} is completed. {\hfill $\square$}

\end{document}